\documentclass[superscriptaddress, reprint, amsmath, amssymb, aps, prb, floatfix]{revtex4-2}

\usepackage{graphicx}
\usepackage{xcolor}
\usepackage{hyperref}
\usepackage{orcidlink}
\hypersetup{colorlinks=true, linkcolor=blue, filecolor=magenta, urlcolor=magenta}

\definecolor{pink}{rgb}{1,0,0.9}
\usepackage{tikz}
\usepackage{placeins}   
\usetikzlibrary{decorations.pathreplacing, arrows.meta}

\newcommand{\Pt}{\hat{P}}

\newcommand{\ceff}{c_{\rm eff}}
\newcommand{\xie}{\xi_{\mathbf 8}}

\newcommand{\one}{\mathbb{I}}   

\begin{document}
\renewcommand{\topfraction}{0.9}\renewcommand{\bottomfraction}{0.8}\renewcommand{\textfraction}{0.07}
\renewcommand{\floatpagefraction}{0.8}\renewcommand{\dbltopfraction}{0.9}\renewcommand{\dblfloatpagefraction}{0.8}
\setcounter{topnumber}{3}\setcounter{bottomnumber}{2}\setcounter{totalnumber}{5}\setcounter{dbltopnumber}{2}

\title{
Study of the frustrated SU(3)
Heisenberg chain with SU(3)-symmetric matrix product states}

\author{O.~R.~Hryniv}
\affiliation{Akhiezer Institute for Theoretical Physics, NSC KIPT,
  Akademichna 1, 61108 Kharkiv, Ukraine}

\author{I.~V.~Lukin}
\affiliation{Akhiezer Institute for Theoretical Physics, NSC KIPT,
  Akademichna 1, 61108 Kharkiv, Ukraine}
\affiliation{Haiqu, Inc., 95 Third Street, San Francisco, CA 94103, USA}

\author{A.~G.~Sotnikov}
\affiliation{Akhiezer Institute for Theoretical Physics, NSC KIPT,
  Akademichna 1, 61108 Kharkiv, Ukraine}
\affiliation{Karazin Kharkiv National University,
  Svobody Square 4, 61022 Kharkiv, Ukraine}

\begin{abstract}
We study the SU(3) Heisenberg chain with nearest- and next-nearest-neighbor antiferromagnetic exchange $J_1$ and
$J_2$. Field theory predicts that the
trimerized valence-bond solid of this chain has a finite extent in $\alpha=J_2/J_1$, followed by a gapless SU(3)$_1$
phase, but earlier numerical estimates of the upper boundary range from $\alpha\approx2.1$ (exact diagonalization) to $\alpha\approx3.8$ (iDMRG). 
We use the variational uniform matrix product state (VUMPS) algorithm with the full non-Abelian SU(3) symmetry imposed on the virtual bonds, reaching bond dimensions up to $D\approx8000$, and locate the phase boundaries by finite-entanglement scaling. 
Resolving the transfer-matrix spectrum and the entanglement spectrum by SU(3) irreducible representation, we find that the trimerized phase extends to $\alpha_{c2}\gtrsim3.5$ and end below $\alpha\approx5$, excluding a boundary near $\alpha\approx2$. 
At larger $\alpha$ our numerical analysis of the effective central charge is consistent with the predicted massless flow from SU(3)$_2$ to SU(3)$_1$.
\end{abstract}

\maketitle

\section{Introduction}\label{sec:intro}

Lattice models with SU($N$) symmetry describe Mott insulators of alkaline-earth-like atoms with $N$
populated nuclear-spin states~\cite{GorshkovNatPhys2010,CazalillaRey2014}, which have been realized with
$^{173}$Yb and $^{87}$Sr in optical lattices~\cite{Taie2012,Pagano2014,Zhang2014,Hofrichter2016}.
For $N>2$ two spins in the fundamental representation cannot form a singlet; the smallest singlet involves
$N$ sites. Frustration therefore favors $N$-site valence-bond solids rather than dimers.

The simplest frustrated model of this kind is the $J_1$--$J_2$ SU(3) Heisenberg chain [Eq.~(\ref{eq:H})].
At zero next-nearest-neighbor coupling ($J_2=0$), it is the integrable Sutherland chain~\cite{Sutherland1975}, a gapless SU(3)$_1$ Wess--Zumino--Witten (WZW) critical point with central charge $c=2$. Increasing $J_2$ drives a marginal current--current coupling through zero and opens a gap into a threefold-degenerate trimerized phase at
$\alpha_{c1}\approx0.48$~\cite{Corboz2007trimer,Rachel2009,HCL2023}. The fate of this phase at large
$\alpha$ was clarified by Herviou, Capponi and Lecheminant (HCL)~\cite{HCL2023}: viewing the chain as a
zigzag ladder of two SU(3) chains coupled by $J_1$, they showed that for odd $N$, the inter-chain coupling
gaps a $\mathbb{Z}_N$ parafermion sector and leaves an SU($N$)$_2$ theory perturbed by its adjoint field,
which flows without a gap to SU($N$)$_1$ \cite{Lecheminant2015}. The trimerized phase thus has a finite extent, and a second
gapless SU(3)$_1$ phase appears at large~$\alpha$.

The location of the upper boundary $\alpha_{c2}$ for the trimerized phase and the nature of the large-$\alpha$ phase are not settled numerically. Exact diagonalization (ED) on rings of up to 27 sites gives $\alpha_{c2}=2.1(1)$ with strong finite-size effects, while iDMRG with U(1)$\times$U(1) color conservation and bond dimension
$\chi\le8192$ shows trimerization up to $\alpha\approx3.8$, beyond which the data jump at a $\chi$-dependent
coupling~\cite{HCL2023}; an earlier study placed the boundary near $3.5$~\cite{Corboz2007trimer}. At large
$\alpha$, HCL obtained an effective central charge $c\simeq3.2$, close to the SU(3)$_2$ value $16/5$, and
argued that full SU(3) symmetry would be needed to resolve the flow to $c=2$, in analogy with the SU(3)
chain in the symmetric representation~\cite{Nataf2016,NatafGozelMila2021}.

Here we address both questions with variational uniform matrix product states (VUMPS)
\cite{ZaunerStauber2018,Vanderstraeten2019} with exact SU(3) symmetry on the virtual bonds, at bond
dimensions up to $D\approx8000$ [about 540 SU(3) multiplets]. Our main results are:
\begin{enumerate}
\item An injective SU(3)-symmetric MPS with the fundamental representation on every site cannot be
invariant under one-site translation; finite-$D$ states therefore break translation and generically show some trimerization, so that the phase boundaries have to be located by finite-entanglement scaling
(Sec.~\ref{sec:model}).
\item At the lower transition, the extrapolated trimerization becomes positive at $\alpha_{c1}\approx0.48$, once states up to $D=4000$ are employed, consistent with the earlier ED estimate. By contrast, a fit of the essential-singularity form expected for this SU(3)-symmetric marginal transition cannot locate $\alpha_{c1}$ at reachable bond dimensions (Sec.~\ref{sec:lower}).
\item From the bond-dimension scaling of the octet correlation length, the entanglement entropy and the trimerization, the trimerized phase extends to $\alpha_{c2}\gtrsim3.5$; where it ends is not resolved at the bond dimensions reached (Sec.~\ref{sec:upper}).
\item Above $\alpha_{c2}$, for $\alpha\in[4,10]$, the effective central charge lies below $16/5$, and where the crossover length is shortest (moderate $\alpha\approx4$), it keeps decreasing with $D$ (Sec.~\ref{sec:large}).
\item The jumps of the iDMRG data of Ref.~\cite{HCL2023} are crossings, at fixed bond dimension, to states outside the SU(3)-symmetric ansatz, which we reproduce with an Abelian iDMRG approach; no SU(3)-symmetric state of either entanglement structure reproduces them (Sec.~\ref{sec:jumps}).
\end{enumerate}
After the model and the method (Sec.~\ref{sec:model}) we give an overview of the phase diagram
(Sec.~\ref{sec:overview}), determine the lower and the upper boundary of the trimerized phase
(Secs.~\ref{sec:lower} and \ref{sec:upper}), discuss the large-$\alpha$ regime (Sec.~\ref{sec:large}), and compare
with previous work (Sec.~\ref{sec:jumps}).

\section{Model and method}\label{sec:model}

\subsection{Model}

\begin{figure*}[!tbp]
\centering
\begin{tikzpicture}[scale=1.4]
  \colorlet{colR}{red!70!black}
  \colorlet{colG}{green!55!black}
  \colorlet{colB}{blue!65!black}
  \def\rsit{0.215}  

  \foreach \xa/\xb in {0/2, 2/4, 4/6}{
    \draw[gray, dashed, semithick] (\xa,0) -- (\xb,0);
  }
  \foreach \xa/\xb in {1/3, 3/5}{
    \draw[gray, dashed, semithick] (\xa,1.55) -- (\xb,1.55);
  }

  \foreach \xb/\xt in {0/1, 2/3, 4/5}{
    \draw[black, thick] (\xb,0) -- (\xt,1.55);
  }
  \foreach \xt/\xb in {1/2, 3/4, 5/6}{
    \draw[black, thick] (\xt,1.55) -- (\xb,0);
  }

  \node[font=\normalsize] at (-0.85,0)    {$\cdots$};
  \node[font=\normalsize] at (-0.6,1.55)  {$\cdots$};
  \node[font=\normalsize] at ( 6.85,0)    {$\cdots$};
  \node[font=\normalsize] at ( 6.6,1.55)  {$\cdots$};

  \filldraw[colR, draw=black, line width=0.5pt] (0,0) circle (\rsit);
  \filldraw[colG, draw=black, line width=0.5pt] (2,0) circle (\rsit);
  \filldraw[colB, draw=black, line width=0.5pt] (4,0) circle (\rsit);
  \filldraw[colR, draw=black, line width=0.5pt] (6,0) circle (\rsit);
  \filldraw[colG, draw=black, line width=0.5pt] (1,1.55) circle (\rsit);
  \filldraw[colB, draw=black, line width=0.5pt] (3,1.55) circle (\rsit);
  \filldraw[colR, draw=black, line width=0.5pt] (5,1.55) circle (\rsit);

  \node[above left,  font=\scriptsize] at (0.35,0.88) {$J_1$};
  \node[above right, font=\scriptsize] at (1.65,0.88) {$J_1$};
  \node[below,       font=\scriptsize] at (1.0, 0)    {$J_2$};
  \node[above,       font=\scriptsize] at (2.0, 1.55) {$J_2$};

  \draw[-{Stealth[length=4pt]}, gray, thin]
    (0.32,0.62) -- (-0.55,0.62)
    node[left, font=\scriptsize, align=right, text=gray]
      {``rung''\\[-1pt]$E_{\rm rung}$};
  \draw[-{Stealth[length=4pt]}, gray, thin]
    (1.68,0.62) -- (2.65,0.62)
    node[right, font=\scriptsize, align=left, text=gray]
      {``diag''\\[-1pt]$E_{\rm diag}$};

  \draw[decorate,
        decoration={brace, mirror, amplitude=5pt},
        gray]
    (-0.28,-0.52) -- (5.28,-0.52);
  \node[below, font=\scriptsize, text=gray] at (2.5,-0.72)
    {unit cell \;($N_{\rm UC}=6$)};

  \begin{scope}[xshift=8.6cm, yshift=0.2cm]
    \draw[rounded corners=3pt, gray!40, fill=gray!6, line width=0.6pt]
      (-0.5,-0.5) rectangle (2.5,2.6);
    \node[font=\footnotesize, text=black] at (1.0,2.25)
      {$\hat{P}_{ij}$};

    \filldraw[colR, draw=black, line width=0.5pt] (0.35,1.6) circle (0.22);
    \filldraw[colG, draw=black, line width=0.5pt] (1.65,1.6) circle (0.22);
    \node[font=\scriptsize] at (0.35,1.6) {\textbf{1}};
    \node[font=\scriptsize, text=white] at (1.65,1.6) {\textbf{2}};
    \node[below, font=\scriptsize] at (0.35,1.36) {$i$};
    \node[below, font=\scriptsize] at (1.65,1.36) {$j$};

    \draw[-{Stealth[length=5pt]}, thick, gray!70]
      (1.0,1.22) -- (1.0,0.78);

    \filldraw[colG, draw=black, line width=0.5pt] (0.35,0.42) circle (0.22);
    \filldraw[colR, draw=black, line width=0.5pt] (1.65,0.42) circle (0.22);
    \node[font=\scriptsize] at (0.35,0.42) {\textbf{2}};
    \node[font=\scriptsize, text=white] at (1.65,0.42) {\textbf{1}};
  \end{scope}
\end{tikzpicture}
\caption{%
    \textbf{Zigzag chain geometry and the SU(3) permutation operator.}
    Sites (circles) carry one of three colors (red~=~1, green~=~2, blue~=~3), illustrating the period-3 pattern of the trimerized phase.
    Solid lines are nearest-neighbor ($J_1$) bonds; within the zigzag, two geometrically distinct subtypes arise -- ``rung'' bonds (bottom$\to$top, carrying energy $E_{\rm rung}$) and ``diagonal'' bonds (top$\to$next-bottom, $E_{\rm diag}$) -- which can have different expectation values whenever translational symmetry is broken.
    Dashed lines are next-nearest-neighbor ($J_2=\alpha J_1$) bonds running along each row
    ($E_{J_2}$), which introduce geometric frustration.
    The bracketed region is one unit cell ($N_{\rm UC}=6$ sites).
    \emph{Inset}: the permutation operator $\hat{P}_{ij}$ exchanges the SU(3) colors of
    sites $i$ and $j$. It plays the same role as $2\mathbf{S}_i\cdot\mathbf{S}_j$ in
    SU(2) but has no two-site singlet eigenstate, so color-neutral clusters require
    three sites.}
\label{fig:model}
\end{figure*}
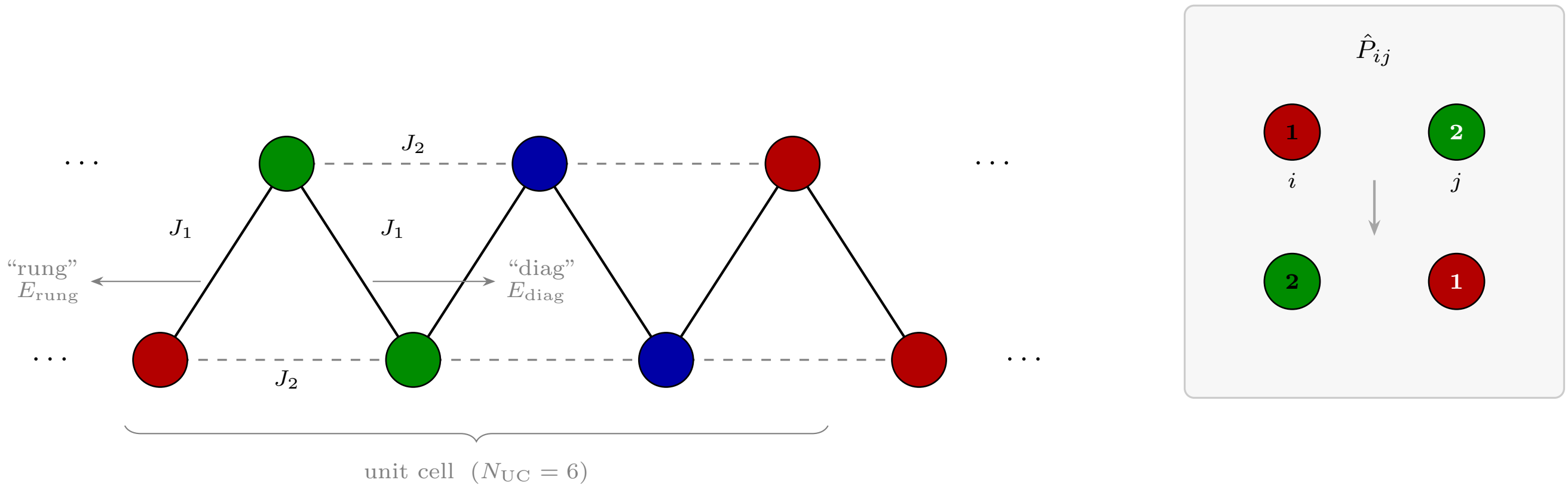

We consider
\begin{equation}
  \hat{H} = J_1\sum_i \Pt_{i,i+1} + J_2\sum_i \Pt_{i,i+2}, \qquad \alpha\equiv J_2/J_1,
  \label{eq:H}
\end{equation}
where each site carries the fundamental representation~$\mathbf 3$ of SU(3) and $\Pt_{ij}$ exchanges the
states of sites $i$ and $j$. With SU(3) generators normalized as ${\rm Tr}(\hat{S}^A\hat{S}^B)=\delta^{AB}/2$ one has
$2\sum_A \hat{S}^A_i\hat{S}^A_j=\Pt_{ij}-1/3$, so Eq.~(\ref{eq:H}) differs from the spin form used in
Ref.~\cite{HCL2023} only by a factor of two, and $\alpha$ coincides with $J_2/J_1$. We set
$J_1=1$. Drawing the chain as a zigzag ladder (Fig.~\ref{fig:model}), the $J_2$ bonds form two legs and the
$J_1$ bonds alternate between ``rung'' and ``diagonal'' bonds. Two sites cannot form an SU(3) singlet,
since $\mathbf 3\otimes\mathbf 3=\mathbf 6\oplus\bar{\mathbf 3}$; the smallest singlet involves three sites,
which is why frustration favors period-3 (trimer) rather than bipartite order.

At $\alpha=0$, Eq.~(\ref{eq:H}) is the Sutherland chain, with exact ground-state energy per site
$e_0=1-\ln3-\pi\sqrt3/9=-0.703212$~\cite{Sutherland1975} and an SU(3)$_1$ critical point with $c=2$. At
$J_1=0$ the two legs decouple into two such chains ($c=4$). For any finite $\alpha$, the inter-leg coupling is relevant: it gaps a $\mathbb Z_3$ parafermion sector ($c=4/5$)
and leaves SU(3)$_2$ ($c=16/5$) perturbed by its adjoint primary, which flows without a gap to
SU(3)$_1$ ($c=2$)~\cite{HCL2023, Lecheminant2015}. Here $c=k(N^2-1)/(k+N)$ for SU($N$)$_k$, i.e. $8/4=2$ for $k=1$ and $16/5$ for $k=2$; the parafermion sector accounts for $4-16/5=4/5$.

We use a unit cell of six sites (three rungs), commensurate with both the period-3 order and the two-leg
structure, as in Ref.~\cite{HCL2023}. All states we obtain have a period of three sites (the bond energies repeat after three sites), thus a three-site cell would describe the same states; the six-site cell, in addition, admits period-2 and
period-6 patterns, which do not occur.

\subsection{SU(3)-symmetric uniform MPS}

Ground states are obtained with the variational uniform matrix-product-state algorithm
(VUMPS)~\cite{ZaunerStauber2018,Vanderstraeten2019}, with the SU(3) symmetry imposed exactly on the virtual
spaces~\cite{MPSKit,TensorKit2025,SUNRepresentations,Alex2011}, as in non-Abelian symmetric DMRG and tensor-network
methods~\cite{McCullochGulacsi2002,Weichselbaum2012}: every tensor is stored as reduced blocks between multiplets,
with the Clebsch--Gordan structure fixed by the symmetry. Each virtual space is a direct
sum of SU(3) irreducible representations (irreps) with multiplicities; the bond dimension $D$ counts states,
so an irrep $\lambda$ with multiplicity $m_\lambda$ contributes $m_\lambda\dim\lambda$. At $D\approx6000$ the
states carry about 420 multiplets ($\approx14$ states per multiplet), and the cost is set by the multiplet
structure rather than by $D$. Because each site adds one box to the Young diagram, the triality
(number of boxes mod 3) of the virtual irreps advances by one from bond to bond: every bond carries irreps of a
single triality class, $t=0$ ($\mathbf 1,\mathbf 8,\mathbf{10},\overline{\mathbf{10}},\mathbf{27},\dots$),
$t=1$ ($\mathbf 3,\bar{\mathbf 6},\mathbf{15},\dots$) or $t=2$ (their conjugates), cycling through the unit cell.
The irreps and their multiplicities are not fixed in advance: they are selected by the singular values, and
irreps up to dimension 64 occur in our states. The truncation keeps or discards whole multiplets, ranked by their
Schmidt values, so the state is exactly SU(3)-symmetric at every step; the triality of the irreps on each bond is
fixed by the fusion rules and is not chosen by hand.

The bond dimension is increased in steps of a factor $1.2$. Each step starts with a two-site infinite DMRG sweep
that truncates to the new $D$; this is the step in which the bond spaces are chosen, i.e., which irreps occur on each bond and with which multiplicities, by keeping the multiplets with the largest Schmidt values. VUMPS then optimizes the tensors at fixed bond spaces until the gradient norm
\begin{equation}
  \delta=\max_{i}\big\|(\one-A_L^iA_L^{i\dagger})\,\tilde A_C^i\big\|,\quad
  \tilde A_C^i=H_{A_C}^iA_C^i/\|H_{A_C}^iA_C^i\| ,
  \label{eq:delta}
\end{equation}
falls below $10^{-9}$ (at most 1000 iterations). Here $A_L^i$ and $A_C^i$ are the left-orthonormal and center-site tensors of site~$i$ of the cell, and $H_{A_C}^i$ is the effective Hamiltonian acting on $A_C^i$; $\delta$ is the norm of the energy gradient projected onto the tangent space of the MPS manifold~\cite{ZaunerStauber2018, Vanderstraeten2019} and vanishes at a variational optimum.

\subsection{No single-site translation invariance at finite $D$}\label{sec:noTI}

The fundamental representation is a projective representation of PSU(3)$=$SU(3)$/\mathbb Z_3$: the center
element $z=\omega\one$, $\omega=e^{2\pi i/3}$, acts as the phase $\omega$ on each site. Consider a single-site
uniform MPS $A^s$ with this symmetry. Each site adds one box to the Young diagram, so $A^s$ maps virtual irreps of
triality $t$ to triality $t+1$; with $Z=\omega^{t}$ on the triality-$t$ blocks of the virtual space,
$ZA^sZ^{-1}=\omega A^s$. If $\rho$ is the fixed point of the transfer map $\mathcal E(X)=\sum_sA^sXA^{s\dagger}$, then $\mathcal E(Z\rho)=\omega^{-1}Z\rho$: since $Z$ is invertible, $Z\rho\neq0$, and $\omega^{-1}\neq1$ is a second eigenvalue of modulus one. An injective MPS has a unique eigenvalue of modulus one, so an injective
SU(3)-symmetric MPS cannot be invariant under one-site translation. This is the matrix-product-state form of the
Lieb--Schultz--Mattis theorem \cite{LiebSchultzMattis1961, AffleckLieb1986}, shown for SU(2) with half-integer spin in Ref.~\cite{Sanz2009}; a complete derivation is given in Appendix~\ref{app:proof}. 

A translation-invariant SU(3)-symmetric MPS exists only in non-injective form, as the equal-weight superposition of the three translates of a period-3 injective state; its correlations do not cluster and it carries no information beyond its components. 
In the SU(3)-symmetric representation, the translation breaking of an injective state is visible directly in the entanglement spectrum. On cuts with $t\neq0$ every irrep has dimension divisible by three, so every Schmidt value is at least threefold degenerate, while the $t=0$ cuts contain the singlet sector with non-degenerate values. The spectra of neighboring cuts therefore differ, and every finite-$D$ state breaks single-site translation: the bond energies are modulated with a period of three sites, with $T>0$ in all our states, also in a gapless phase (at $\alpha=0$, $T=0.006$ at $D=1034$). The distinction between the gapless and the trimerized phases is then a matter of $D$-scaling: $T(D)$, $\xie(D)$ and $S(D)$ saturate in the trimerized phase, while in a gapless phase $T\to0$ and $\xie$, $S$ grow without bound.

\subsection{Observables}

\emph{Correlation length.} The transfer matrix of the six-site cell, $\mathcal T=\sum_{\mathbf s}A^{\mathbf s}\otimes
\bar A^{\mathbf s}$, is block diagonal in the SU(3) irrep $\lambda$ of its channel (only triality-zero irreps occur).
Let $t_{\lambda,j}$ be its eigenvalues in sector $\lambda$, ordered by modulus. For a normalized injective MPS the
largest eigenvalue is $t_{\mathbf 1,1}=1$, in the singlet sector; it corresponds to the norm of the state and carries
no correlations. We therefore define, in units of sites,
\begin{equation}
  \xi_\lambda=-\frac{6}{\ln|t_{\lambda,1}|}\quad(\lambda\neq\mathbf 1),\qquad
  \xi_{\mathbf 1}=-\frac{6}{\ln|t_{\mathbf 1,2}|} .
\end{equation}
The adjoint length $\xie$ is the longest at all couplings, as also found in Ref.~\cite{HCL2023}.

\emph{Entanglement.} Cutting the infinite chain at one of the six bonds of the unit cell divides it into two
half-infinite parts. The reduced density matrix of one part, $\rho=\mathrm{Tr}_{\rm R}|\psi\rangle\langle\psi|$, has
the squared Schmidt values $s_k^2$ of that bond as eigenvalues, which follow from the bond matrix of the MPS in mixed
canonical form. With SU(3) symmetry they come in multiplets: a Schmidt value in irrep $\lambda$ is
$\dim\lambda$-fold degenerate. For each cut we compute the entanglement entropy $S=-\sum_ks_k^2\ln s_k^2$ and the
entanglement energies $\epsilon_k=-\ln s_k^2$; we report $S$ averaged over the six cuts, and the largest eigenvalue
$\lambda_{\max}$ of $\rho$ over the cuts together with the irrep that carries it.

\emph{Trimerization.} With $E_b=\langle\Pt_{i,i+1}\rangle$ on the nearest-neighbor bonds, the trimerization is
$T=E_{\rm weak}-\frac13\sum_{k=0}^{2}E_{b+k}$ on the weak bond~\cite{HCL2023}. In a period-3 state with
bond energies $\{s,s,w\}$, $T=2(w-s)/3=(2/\sqrt3)\,\sigma_{\rm rung}$, where
\begin{equation}
  \sigma_{\rm rung}=\Big[\frac12\sum_{k=1}^{3}\big(E^{(k)}_{\rm rung}-\bar E_{\rm rung}\big)^2\Big]^{1/2}
\end{equation}
is the sample standard deviation of the three rung energies $E^{(k)}_{\rm rung}$ of the cell, with mean
$\bar E_{\rm rung}=\frac13\sum_kE^{(k)}_{\rm rung}$; for $\{s,s,w\}$ it equals $|w-s|/\sqrt3$.

\emph{Finite-entanglement scaling.} At a critical point a finite-$D$ MPS has a finite $\xi(D)\propto D^\kappa$,
with $\kappa=6/[c(\sqrt{12/c}+1)]$~\cite{Pollmann2009}, and $S=(c/6)\ln\xi+{\rm const}$~\cite{Calabrese2004}.
We extract an effective central charge $\ceff(D)=6\,dS/d\ln\xie$ from the slope of $S$ versus $\ln\xie$ over the three
most recent states up to and including $D$ (two states whose $D$ differ by less than $10\%$, which occur where a run was resumed, are merged); in a gapped phase $\ceff\to0$ as $\xie$ saturates.

\emph{Extrapolation of the trimerization.} Every finite-$D$ SU(3)-symmetric state has $T>0$, also in a gapless
phase (see Sec.~\ref{sec:noTI}). Therefore, the value of $T$ at a given bond dimension does not indicate whether the chain is trimerized; the question is whether $T(D)$ tends to zero or to a finite value as $D\to\infty$. A finite-$D$ MPS describes the infinite system up to the length scale $\xie(D)$, which plays the role of a finite system size. In a gapless phase an order parameter that vanishes in the thermodynamic limit then decays as $T(D)\propto\xie^{-\Delta}$ \cite{Tagliacozzo2008, Pirvu2012}, where $\Delta$ is the scaling dimension of the corresponding operator \cite{Affleck1986, Affleck1988}. For the trimerization in the SU(3)$_1$ phase this is the operator ${\rm Tr}\,g$, built from the primary field $g$ in $(\mathbf3,\bar{\mathbf3})$ with $h=\bar h=C_2(\mathbf 3)/(k+N)=(4/3)/4=1/3$, so that $\Delta=2h=2/3$. In a gapped phase $T(D)$ instead approaches a finite limit once $\xie(D)$ exceeds the correlation length of the ground state. We therefore fit
\begin{equation}
  T(D)=T_\infty+A\,\xie^{-\Delta}
  \label{eq:Textrap}
\end{equation}
to the states in the larger-$D$ half of each series (at least four), with $\Delta=2/3$ unless stated otherwise. In the gapless phase this form describes the leading finite-$D$ behavior, with $T_\infty=0$ up to logarithmic corrections
near the Sutherland point (Sec.~\ref{sec:lower}); in the trimerized phase it is an interpolation between the critical behavior at short scales and the finite limit at long scales. Since $\xie$ stays
finite, $T_\infty$ is not a measurement of the order parameter at $D=\infty$ but a comparative criterion: we compare it with the values obtained in the same way at gapless reference points ($\alpha=\{0, 5, 6\}$, as suggested by field-theory arguments), which measure the bias of the extrapolation and test its dependence on $\Delta$ by varying the exponent between $1/2$ and $1$.

\subsection{Convergence and bond dimensions}

Of the 315 states used in this work, 278 reach $\delta<10^{-9}$ and 295 reach $\delta<10^{-8}$. States at $\alpha=2.75$--$10$ and $D\gtrsim3500$ that reached the iteration limit were continued at fixed $D$ for up to 1500 further iterations; this changed the energy by less than $10^{-9}$ per site, $S$ by at most $1.2\times10^{-5}$ and $\xie$ by at most $0.01$ sites. We reach $D=4000$ for $\alpha=0.45$--$0.50$, $D=6000$ for $\alpha=0$, $1$--$2.75$, $6$--$10$ and $20$--$32$, and $D\approx8000$ for $\alpha=3.0$--$4.0$ and $5$; other couplings are taken to $D\approx1000$--$2000$.

\emph{Benchmark.} At $\alpha=0$ and $D=5982$, $e_0=-0.7032119$, $3\times10^{-5}\,\%$ above the exact value.
$\xie$ grows as $D^{0.78}$ (prediction for $c=2$: $\kappa=0.87$) and $\ceff=1.92$, $1.95$, $1.95$, $1.98$ in the
windows $D\in[200,500)$, $[500,1000)$, $[1000,2000)$, $[2000,4000)$, approaching $c=2$ from below. Both the reduced
exponent and the slow approach of $\ceff$ to $c=2$ are expected from the logarithmic corrections generated by the
marginally irrelevant coupling of the Sutherland chain~\cite{AffleckGepnerSchulzZiman1989, Eggert1996, ItoiKato1997, HCL2023}.

\section{Overview of the phase diagram}\label{sec:overview}

\begin{figure}[!htbp]
\centering
\includegraphics[width=\columnwidth]{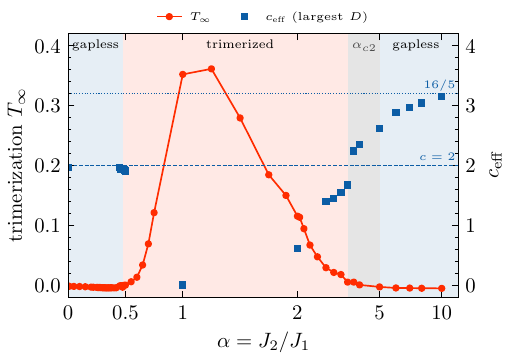}
\caption{Phase diagram along $\alpha=J_2/J_1$ (linear scale at $\alpha<1$ and logarithmic above). Shaded: gapless SU(3)$_1$ phase ($\alpha<\alpha_{c1}=0.48$), trimerized phase, and gapless phase at large $\alpha$, where the SU(3)$_2\to$SU(3)$_1$ crossover is expected; grey band: $\alpha=3.5$--$5$, where the upper boundary lies but is not resolved ($\alpha_{c2}\gtrsim3.5$, Sec.~\ref{sec:upper}). Red: trimerization extrapolated to $D\to\infty$. Blue: effective central charge at the largest bond dimension; dashed and dotted lines mark $c=2$ and $c=16/5$.}
\label{fig:phase}
\end{figure}

Figure~\ref{fig:phase} summarizes the phase diagram. At small~$\alpha$ the chain is gapless, as at the Sutherland point; above a lower transition $\alpha_{c1}$ it is trimerized, a threefold-degenerate period-3 valence-bond solid whose order parameter is largest near $\alpha\approx1.25$; above an upper boundary $\alpha_{c2}$ it is gapless again, and at large $\alpha$ the effective central charge follows the crossover from two decoupled chains towards SU(3)$_1$. The two boundaries are determined in Secs.~\ref{sec:lower} and \ref{sec:upper}, and the large-$\alpha$ regime is discussed in Sec.~\ref{sec:large}. Here, we describe the ground state in the three regions; the
trimerization, the largest Schmidt weight, and the inverse octet correlation length are shown versus $\alpha$ for several bond dimensions in Fig.~\ref{fig:hclstyle}.

\begin{figure}[!htbp]
\centering
\includegraphics[width=\columnwidth]{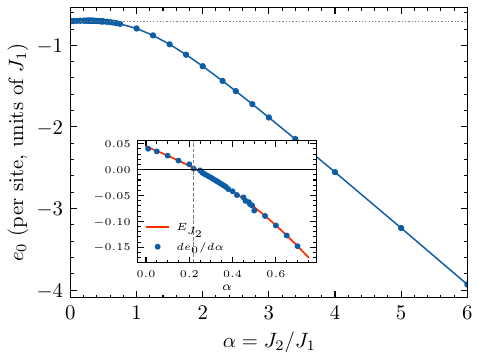}
\caption{Ground-state energy per site. Dotted line: exact value at $\alpha=0$~\cite{Sutherland1975}.
Inset: $de_0/d\alpha$ from finite differences (points) and the measured energy of the $J_2$ bonds (line), equal by
the Hellmann--Feynman theorem; the dashed line marks the maximum of $e_0$, where $E_{J_2}$ changes sign. States
at $D\approx1000$ (inset) and the largest $D$ (main panel).}
\label{fig:energy}
\end{figure}

\begin{figure}[!htbp]
\centering
\includegraphics[width=0.9\columnwidth]{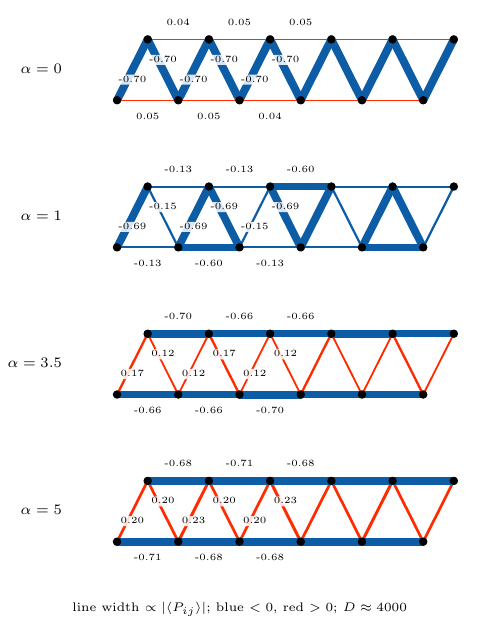}
\caption{Bond energies $\langle\Pt_{ij}\rangle$ over two unit cells at $D\approx4000$. Line width is proportional
to $|\langle\Pt_{ij}\rangle|$; blue: negative, red: positive. Numbers give the bond energies of the first unit cell;
the pattern repeats in the second.}
\label{fig:bonds}
\end{figure}

\emph{Energy.} The ground-state energy per site (Fig.~\ref{fig:energy}) is smooth over the whole range. Since
$e_0=E_{J_1}+\alpha E_{J_2}$ with $E_{J_2}=\langle\Pt_{i,i+2}\rangle$, the Hellmann--Feynman theorem gives
$de_0/d\alpha=E_{J_2}$; numerical derivatives of $e_0$ agree with the measured $E_{J_2}$ to $1.5\times10^{-4}$ (median
deviation at $D\approx1000$; with states up to $D=4000$ near $\alpha_{c1}$, to $7\times10^{-6}$). The energy has a shallow maximum at
$\alpha\approx0.22$--$0.25$, where $E_{J_2}$ changes sign; this maximum is unrelated to the transition at
$\alpha_{c1}$.

\emph{Bond energies.} Figure~\ref{fig:bonds} shows all bond energies of the six-site cell at $D\approx4000$. At
$\alpha=0$ the $J_1$ bonds are uniform within $0.6\%$, a finite-$D$ residual of the translation-symmetry breaking
discussed in Sec.~\ref{sec:noTI}. At $\alpha=1$ the trimers are visible directly: two strong $J_1$ bonds
($\langle\Pt\rangle=-0.69$) and one strong $J_2$ bond ($-0.60$) form a triangle, separated from the next one by weak
bonds ($-0.15$ and $-0.13$). At $\alpha=3.5$ and $5$ the energy resides in the legs ($\langle\Pt\rangle\approx-0.66$ to
$-0.71$) and the $J_1$ bonds are frustrated ($\langle\Pt\rangle>0$); a period-3 modulation remains in both bond
types and decreases with $\alpha$.

\emph{Entanglement spectrum.} Resolved by SU(3) irrep (Appendix~\ref{app:es}), the entanglement spectrum shows the
structure of the trimerized phase directly: at $\alpha=1$ the $t=0$ cut, which separates two trimers, is dominated by
a single singlet level ($\lambda_{\max}=0.75$), so that the state is close to a product of SU(3) singlets on three
sites. The singlet--octet entanglement gap on this cut varies smoothly with $\alpha$ and shows no jump at any $D$. The singlet carries the largest Schmidt value on the $t=0$
cut for all $\alpha\le10$.

\section{The lower transition}\label{sec:lower}

\begin{figure}[!htbp]
\centering
\includegraphics[width=\columnwidth]{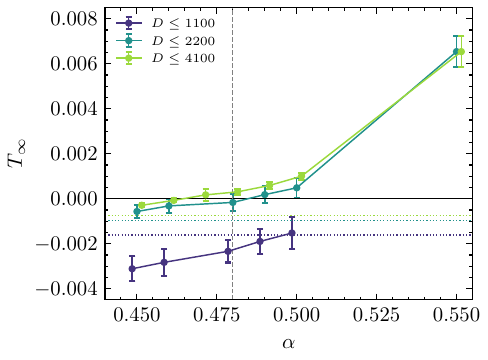}
\caption{Extrapolated trimerization $T_\infty$ near the lower transition, from states with $200\le D\le D_{\rm cut}$.
Dotted lines: $T_\infty$ at the gapless Sutherland point $\alpha=0$ for the same $D_{\rm cut}$, a measure of
the bias of the extrapolation. Dashed line: $\alpha_{c1}=0.48$ from ED level crossings~\cite{HCL2023}.}
\label{fig:lower}
\end{figure}

\begin{figure*}[!tbp]
\centering
\includegraphics[width=\textwidth]{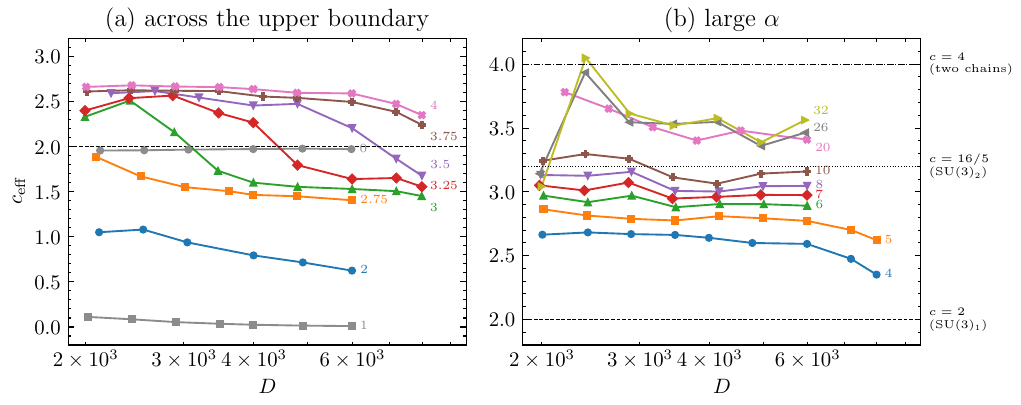}
\caption{Effective central charge versus bond dimension: slope of $S$ versus $\ln\xie$ over the three most recent
states up to $D$, plotted at $D$. Where the main run of a coupling starts above $D\approx1600$, the states below its
first $D$ come from a second run started at $D\approx1400$; at the junction both runs give the same state within the
difference in $D$. (a) Across the upper boundary, with the gapless ($\alpha=0$) and gapped ($\alpha=1$) references. (b) Large $\alpha$, including $\alpha=20$, $26$ and $32$; lines: $c=4$
(two decoupled chains), $16/5$ (SU(3)$_2$) and $2$ (SU(3)$_1$). $\ceff$ probes the
renormalization-group trajectory $c=4\to16/5\to2$ at the length scale $\xie(D)$; since the crossover lengths grow with
$\alpha$, at fixed $D$ it increases with $\alpha$, and the values in (b) are not a plateau at $16/5$
(see Sec.~\ref{sec:large}).}
\label{fig:ceff}
\end{figure*}

At $\alpha_{c1}$ a marginal current--current coupling changes sign~\cite{HCL2023}; the gap opens as
$\exp[-{\rm const}/(\alpha-\alpha_{c1})]$, and just above the transition the correlation length exceeds any length
accessible to finite-entanglement methods. Indeed, at $D=4000$ the states at $\alpha=0.45$--$0.50$ have
$\xie\approx380$--$410$ sites, still growing as $D^{0.75\text{--}0.79}$, and $\ceff=1.98$ for $2000\le D<4000$, which cannot be distinguished
from a gapless state by $\xie$ or $S$ alone.

The extrapolated trimerization $T_\infty$ resolves the transition once the bond dimension is large enough
(Fig.~\ref{fig:lower}). With $D\le1100$ the extrapolation is biased: $T_\infty$ is negative for all
$\alpha\le0.50$ and at the Sutherland point ($-0.0016$). The bias comes from the logarithmic corrections
generated by the marginal coupling, which the power law $\xie^{-2/3}$ does not capture; it decreases with $D$
(at $\alpha=0$: $-0.0016$, $-0.0010$, $-0.0008$ for $D\le1100$, $2200$, $4100$), and it is smallest close to
$\alpha_{c1}$, where the marginal coupling vanishes. Using states up to $D=4000$, $T_\infty$ is zero within
$3\times10^{-4}$ for $\alpha=0.45$--$0.47$ and positive and increasing from $\alpha=0.48$ on: $T_\infty=0.0003(1)$,
$0.0006(2)$, $0.0010(2)$ and $0.0065(7)$ at $\alpha=0.48$, $0.49$, $0.50$ and $0.55$. Here the exponent is not
only assumed: fits with a free exponent give $0.63$--$0.67$ at $\alpha=0.45$--$0.50$, consistent with $\Delta=2/3$.
This gives
$\alpha_{c1}\le0.48$, in agreement with the level-crossing analysis of ED spectra, $\alpha_{c1}\approx0.48$
\cite{HCL2023}, which locates the sign change of the marginal coupling directly. Because the order parameter is
exponentially small just above $\alpha_{c1}$, the data cannot exclude a transition slightly below $0.48$; we
adopt $\alpha_{c1}=0.48$.

A fit of the essential-singularity form expected for this transition does not improve on this
(Appendix~\ref{app:fits}). With SU($N$) symmetry the gap opens as $\exp[-b/(\alpha-\alpha_c)]$, so that
$T_\infty=A\exp[-b'/(\alpha-\alpha_c)]$; fitted to $T_\infty(\alpha)$, this form gives $\alpha_c=0.40$ (statistical
error $\pm0.007$), well below $\alpha_{c1}\approx0.48$, and $\alpha_c$ moves to $0.40$--$0.49$ when a constant offset is
added to absorb the bias of the extrapolation.

\section{Upper boundary of the trimerized phase}\label{sec:upper}

\begin{figure}[!htbp]
\centering
\includegraphics[width=0.85\columnwidth]{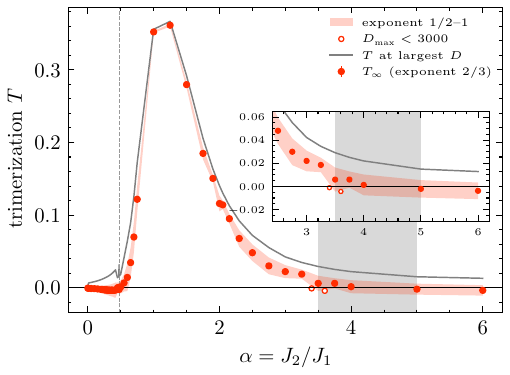}
\caption{Trimerization extrapolated to $D\to\infty$, $T_\infty$ (points, exponent $2/3$), the range of $T_\infty$
for exponents between $1/2$ and $1$ (band), and $T$ at the largest $D$ (line); inset: the region near $\alpha_{c2}$.
Open points: couplings with $D_{\max}<3000$ near the upper boundary ($\alpha=3.4$, $3.6$), not included in the band.
Dashed: $\alpha_{c1}=0.48$; grey band: $\alpha=3.5$--$5$, where the upper boundary lies but is not resolved.}
\label{fig:profile}
\end{figure}

We locate $\alpha_{c2}$ from the bond-dimension dependence of the effective central charge and of the
extrapolated trimerization, calibrated on two reference points (Fig.~\ref{fig:ceff}a). At the Sutherland point
$\ceff=1.96$--$1.98$ for all $D$ from $2000$ to $6000$, as for a gapless state with $c=2$. Deep in the trimerized phase, at $\alpha=1$, $\ceff$ falls to $0.01$ and $\xie$ grows only slowly, to $35$ sites at $D=6000$. A gapped state whose correlation 
exceeds $\xie(D)$ shows $\ceff$ decreasing with $D$, through and below 2; a gapless state in the SU(3)$_1$ phase approaches $c=2$, and in the crossover regime discussed in Sec.~\ref{sec:large} stays above it.

For $\alpha=2$ to $3.5$, $\ceff$ lies below 2 and keeps decreasing up to the largest bond dimensions:
$\ceff=0.63$ ($\alpha=2$), $1.41$ ($2.75$, both at $D=6000$), $1.45$ ($3.0$), $1.56$ ($3.25$) and $1.68$ ($3.5$,
all at $D\approx8000$). At $\alpha=3.75$ and $4.0$ it is $2.24$ and $2.35$ ($D\approx8000$), decreasing slowly. The
curves for $\alpha=3.0$--$3.5$ have kinks, where $\ceff$ drops by up to $0.5$ between consecutive points. They do not come
from the entropy: the two leading eigenvalues of the octet sector of the transfer matrix are nearly degenerate
($|t_{\mathbf8,2}/t_{\mathbf8,1}|>0.97$ for $D\gtrsim1400$) and cross at $D\approx2400$, $4000$ and $5000$--$6000$ for
$\alpha=3.0$, $3.25$ and $3.5$, where $\xie$ switches to the faster-growing branch. With the length
$\xi_{\rm pair}=-6/\ln\sqrt{|t_{\mathbf8,1}t_{\mathbf8,2}|}$, which is smooth through the crossings, $\ceff$ decreases
smoothly with $D$ and reaches $1.61$, $1.66$, $1.78$ and $1.91$ at $\alpha=2.75$, $3.0$, $3.25$ and $3.5$ at the largest
$D$, against $1.97$ at $\alpha=0$; with $\xi_{\mathbf 1}$ the values are lower still. Below 2 for every choice of the
length, $\ceff$ thus indicates a gap up to $\alpha=3.5$. For $\alpha\ge3.75$ the outcome depends on the length used
($2.01$ and $2.09$ at $\alpha=3.75$ and $4.0$ with $\xi_{\rm pair}$, below 2 with $\xi_{\mathbf 1}$), so that $\ceff$ does
not locate the boundary there. Quantities that use no correlation length, the growth rate $dS/d\ln D$ of the entropy
and the decay rate $-d\ln T/d\ln D$ of the trimerization, decrease with $D$ at every $\alpha=2.75$--$10$, by
$15$--$40\%$ over our range of $D$, and change smoothly with $\alpha$; only at $\alpha=2$ do they approach zero
($0.04$ and $0.03$ at $D=6000$).

The extrapolated trimerization (Fig.~\ref{fig:profile}) gives the same
picture: $T_\infty=0.030$, $0.022$ and $0.019$ at $\alpha=2.75$, $3.0$ and $3.25$, $0.006$ at both $\alpha=3.5$ and
$3.75$, and $T_\infty=0.0013$ at $\alpha=4.0$, compatible with zero given the bias of the extrapolation
(the same fit gives $T_\infty=-0.002$ at $\alpha=5$, in the gapless phase). Varying the number of states in the slope from three to five changes $\ceff$ by at most $0.14$ ($0.27$ at $\alpha=3.5$), without moving any value across $c=2$.

Because Eq.~(\ref{eq:Textrap}) with $\Delta=2/3$ describes only the gapless SU(3)$_1$ phase (Sec.~\ref{sec:model}),
Fig.~\ref{fig:profile} also shows $T_\infty$ for exponents between $1/2$ and $1$ (band). Deep in the phase $T_\infty$ hardly depends on the exponent ($0.351$--$0.354$ at $\alpha=1$), and fits with a free
exponent give $1.0$--$1.6$ for $\alpha=2$--$3.5$ and larger values of $T_\infty$, so that $\Delta=2/3$ is the
conservative choice there. At the gapless reference points $\alpha=0$, $5$ and $6$ the extrapolation gives values
between $-0.011$ and $+0.006$ over this range of exponents, which sets the resolution of the criterion $T_\infty>0$.
For every exponent in the range, $T_\infty$ at $\alpha=3.5$ lies above the values at $\alpha=5$ and $6$ (e.g.\ $0.016$
against $0.006$ and $0.003$ for exponent $1$), whereas at $\alpha=4.0$ it lies only slightly above them. Using the singlet correlation length instead of
$\xie$ in the extrapolation shifts $T_\infty$ upward by $0.003$--$0.007$ at all $\alpha=2.75$--$6$, including the gapless reference points, and leaves this comparison unchanged.

We conclude that the trimerized phase extends at least to $\alpha=3.5$: $T_\infty$ lies clearly above the gapless reference values up to $\alpha=3.25$ ($0.02$--$0.03$) and above them for every exponent at $\alpha=3.5$, and $\ceff$ is below 2 for every choice of the correlation length up to $\alpha=3.5$. At $\alpha\ge5$, the data behave as at the gapless reference points; where the trimerized phase ends between $\alpha=3.5$ and $5$ is not resolved at the bond dimensions reached here.

\section{Large $\alpha$: the SU(3)$_2\to$SU(3)$_1$ crossover}\label{sec:large}

For SU($N$)$_k$, $c=k(N^2-1)/(k+N)$: $c=4$ for two decoupled SU(3)$_1$ chains, $16/5$ for SU(3)$_2$ and $2$ for
SU(3)$_1$. Beyond $\alpha_{c2}$ the field theory predicts that the inter-leg coupling first gaps the $\mathbb Z_3$
sector, leaving SU(3)$_2$, which then flows without a gap to SU(3)$_1$~\cite{HCL2023, Lecheminant2015}.

With SU(3) symmetry the effective central charge decreases from the two-chain value towards the SU(3)$_1$ value
(Fig.~\ref{fig:ceff}b). For all $\alpha=4$--$10$ it lies below $16/5$ at $D\gtrsim3000$: $2.35$ ($\alpha=4$) and $2.62$
($5$) at $D\approx8000$, and $2.89$--$3.16$ for $\alpha=6$--$10$ at $D=6000$. At $\alpha=4$ and $5$ it keeps decreasing with $D$ (at $\alpha=5$, from $2.81$ at $D\approx4200$ to $2.62$ at $D\approx8000$), and at every $\alpha=4$--$10$ the
values from states with $D\ge3000$ are lower than from states with $D\le2000$. 

The departure from $16/5$ is seen first at the smallest $\alpha$, as expected for the massless flow, whose crossover length grows as the inter-leg coupling $J_1/J_2$ decreases.

The values in Fig.~\ref{fig:ceff}(b) are therefore set by the accessible bond dimension rather than forming a plateau:
$\ceff$ measures the central charge at the length scale $\xie(D)$, and the crossover lengths grow with $\alpha$, the
$\mathbb Z_3$ gap length as $(J_2/J_1)^{3/2}$~\cite{HCL2023}. At $\alpha=20$--$32$, where $\xie$ reaches $86$--$94$ sites
at $D\approx6000$, $\ceff$ stays at $3.4$--$3.6$, between $16/5$ and $4$, and the crossover is not resolved. Reaching
$c=2$ itself would require bond dimensions well above $D\approx8000$ even at $\alpha=4$ and $5$.

\section{Comparison with previous work}\label{sec:jumps}

\begin{figure*}[!tbp]
\centering
\includegraphics[width=\textwidth]{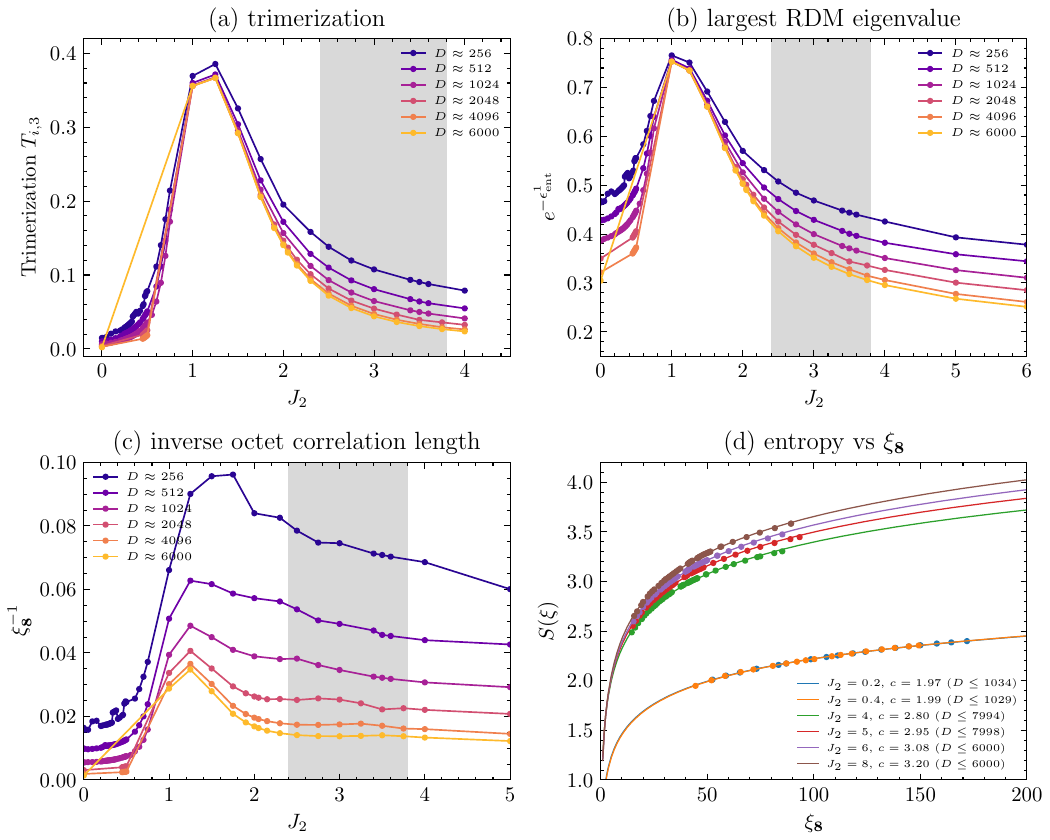}
\caption{SU(3)-symmetric VUMPS data at bond dimensions $D$ matched to the $\chi$ of Ref.~\cite{HCL2023} ($J_1 = 1$, so $J_2=\alpha$).
(a) Trimerization $T$ (cf.\ Fig.~3 of Ref.~\cite{HCL2023}), (b) largest eigenvalue of the half-chain reduced density
matrix (cut between trimers; cf.\ Fig.~4, top), (c) inverse octet correlation length (cf.\ Fig.~5), (d) entanglement entropy versus $\xie$ with
fits $S=(c/6)\ln\xie+b$ over all $D$ (cf.\ Fig.~7). The grey band marks the range $J_2=2.4$--$3.8$ in
which the iDMRG data of Ref.~\cite{HCL2023} jump ($\chi=256$ to $8192$). No jump occurs in the
SU(3)-symmetric data.}
\label{fig:hclstyle}
\end{figure*}

\subsection{Exact diagonalization}

Our bound $\alpha_{c2}\gtrsim3.5$ excludes the ED estimate $\alpha_{c2}=2.1(1)$~\cite{HCL2023}: at $\alpha=2$--$2.5$ all our diagnostics
show a gapped, trimerized state, with $\xie=59$--$71$ sites at $D=6000$ and $T_\infty=0.12$--$0.05$. Since the
correlation length near the upper boundary exceeds 70 sites, rings of up to 27 sites used in ED are far from the
thermodynamic limit there; moreover, level spectroscopy assumes that the transition is described by SU(3)$_1$
with a marginal perturbation, which holds only beyond the crossover length of the SU(3)$_2\to$SU(3)$_1$ flow
(Sec.~\ref{sec:large}). It is compatible with the position $\alpha\approx3.8$ of the jump in the
largest-$\chi$ iDMRG data~\cite{HCL2023}, but is obtained independently of it.

\subsection{iDMRG with Abelian symmetry}

Figure~\ref{fig:hclstyle} shows our data in the same form as the iDMRG data of Ref.~\cite{HCL2023}, at bond
dimensions matched to their $\chi$ (both count states on a bond). For
$\alpha\lesssim2.4$ the two calculations agree well: the inverse octet correlation length at $\alpha=0$ is $0.016$, $0.0095$, $0.0055$ for $D\approx256$, $512$, $1024$ (Ref.~\cite{HCL2023}: $0.019$, $0.010$, $0.006$), the trimerization peaks at $T=0.37$--$0.385$ near $\alpha=1.25$ ($0.38$), the largest
reduced-density-matrix eigenvalue at $\alpha=1$ is $0.77$ ($0.745$ at $J_2=1.2$), and the central charge at
$\alpha=\{0.2,0.4\}$ is $\{1.97, 1.99\}$ (Ref.~\cite{HCL2023}: $\{1.99, 1.98\}$).

In the iDMRG data of Ref.~\cite{HCL2023}, all these quantities jump at a coupling that depends on $\chi$,
from $J_2\approx2.4$ at $\chi=256$ to $\approx3.8$ at $\chi=8192$, after which $T$ is zero. Our data show no
such jump at any $D$ up to $8000$ [Fig.~\ref{fig:hclstyle}(a)--(c) show $D\le6000$; grey band: range of the jumps]. Because an injective SU(3)-symmetric MPS
cannot become translation invariant, the absence of a jump in our data is not by itself evidence against a
transition; the origin of the jumps is examined below and in Appendix~\ref{app:u1}.

At large $\alpha$, Ref.~\cite{HCL2023} found $c\simeq3.26$--$3.33$ for $J_2=4.6$--$8$ with U(1)$\times$U(1) symmetry
and $\chi\le8192$, close to the SU(3)$_2$ value. With SU(3) symmetry and $D\gtrsim3000$ we find $\ceff$ below $16/5$ for
all $\alpha=4$--$10$ and decreasing with $D$ at $\alpha=4$ and $5$ (Sec.~\ref{sec:large}).

\subsection{Origin of the iDMRG jumps}

Could the jumps of Ref.~\cite{HCL2023} be a switch between two different SU(3)-symmetric states? Such states
exist: on the $t=0$ cuts the largest Schmidt value can be carried by the singlet (``singlet type'') or by the octet
(``octet type''), and in our scans the dominant irrep changes from the singlet to the octet near $\alpha\approx22$. We
therefore optimized both types at the same $\alpha$ and $D$ ($\alpha=2.5$--$30$, $D\approx225$, $490$ and $970$),
starting from a state of each type, and compared their energies. Where the iDMRG data jump, the singlet type is
clearly the ground state: for $\alpha\le10$ it is lower at every $D$, by $10^{-3}$--$10^{-2}$ per site, far above the
convergence error. The two types become nearly degenerate only for $\alpha\ge15$, where their energy difference is
$\lesssim10^{-3}$ per site at $D\approx490$ and $970$ and changes sign with $D$. A switch between SU(3)-symmetric states
therefore cannot produce the jumps, and the change of the dominant irrep near $\alpha\approx22$ is a crossing of two
nearly degenerate approximations, not a phase transition.

Since the position of the iDMRG jumps moves with $\chi$, it is not a measurement of a phase boundary; the
boundaries determined in Secs.~\ref{sec:lower} and \ref{sec:upper} from the $D$-scaling of $\xie$, $S$ and $T$ do not
rely on it. Finally, we checked directly that the jumps are reproduced when only Abelian color conservation is imposed, using
an independent infinite-DMRG code (Appendix~\ref{app:u1}). At $\chi=128$ its energy and entanglement entropy agree
with those of the SU(3)-symmetric state with the same number of states up to $J_2=3.15$ (within $1.1\times10^{-4}$ per
site and $0.01$), and the entropy drops between $J_2=3.15$ and $3.3$; at $J_2=3.5$ the Abelian state is less entangled
($S=2.074$ against $2.167$) and lower in energy by $2.5\times10^{-3}$ per site. 

We also test the symmetry from the Schmidt values and the color charges of the Schmidt states: in an SU(3)-symmetric state the Schmidt values form degenerate multiplets, and the summed Schmidt weight at each color weight is invariant under permutations of the three colors. 
Before the jump, the Abelian states are close in energy but already partially breaking SU(3)-symmetry --- the effect decreases with $\chi$: at $J_2=2.0$ the eight Schmidt weights following the dominant singlet, which carry the color charges of an octet, have a relative spread (max$-$min)/mean of $0.41$, $0.14$ and $0.02$ at
$\chi=32$, $64$ and $128$, and at $\chi=128$ the color-permutation asymmetry of the Schmidt weights is at most $0.022$ for $J_2\le3.15$. After the jump, at $J_2=3.5$ and $\chi=128$, the asymmetry is $0.37$ and the octet weights spread by twice their mean (Table~\ref{tab:u1}): the state breaks SU(3). The same comparison at $\chi=64$ shows the motion of the jump: at $J_2=3.0$ and $3.15$ the $\chi=64$ states lie $(1.9$--$2.3)\times10^{-3}$ below the SU(3)-symmetric ones,
with strongly split octet weights, whereas at $\chi=128$ the same couplings are again close in energy to the SU(3)-symmetric branch.
The broken state is not the translation-invariant superposition of three shifted SU(3)-symmetric states either: each
component would carry only about $\chi/3$ states, giving an energy close to that of an SU(3)-symmetric state with
$D\approx\chi/3$, well above the SU(3)-symmetric state with $D\approx\chi$, and an entropy larger by $\ln3$. 
A continuous symmetry with a non-conserved order parameter cannot be broken spontaneously in the ground state of a one-dimensional chain~\cite{Coleman1973}, and the advantage of the broken state is indeed confined to small bond dimension: at $J_2=3.5$ its energy at $\chi=128$, $-2.2092$, lies above that of the SU(3)-symmetric state at $D=7998$, $-2.2167$. The jump is thus a crossing, at fixed bond dimension, to an SU(3)-breaking state that is favorable only at small $\chi$, and its position moves to larger $J_2$ with increasing $\chi$, as in Ref.~\cite{HCL2023}.

\FloatBarrier
\section{Conclusions}\label{sec:concl}

Using uniform matrix product states with exact SU(3) symmetry at bond dimensions up to $D\approx8000$, we have determined the phase diagram of the SU(3) $J_1$--$J_2$ Heisenberg chain (Fig.~\ref{fig:phase}). A gapless SU(3)$_1$ phase at small $\alpha$ is followed, from $\alpha_{c1}\approx0.48$, by a threefold-degenerate trimerized phase, whose order parameter and correlation length are largest and shortest, respectively, near $\alpha\approx1.25$ ($T_\infty=0.36$, $\xie\approx29$ sites). The trimerized phase extends at
least to $\alpha_{c2}\gtrsim3.5$, and at $\alpha\ge5$ the chain is gapless; where between the two it ends is not resolved at the bond dimensions reached here. This resolves the disagreement between
exact diagonalization ($\alpha_{c2}=2.1$) and iDMRG ($\alpha\approx3.8$) in favor of the larger value, which we obtain from the bond-dimension scaling of the entanglement and of the trimerization rather than from the position of a jump.

The jumps found in iDMRG with U(1)$\times$U(1) symmetry are crossings, at fixed bond dimension, to SU(3)-breaking
states that are favorable only at small bond dimension. An injective SU(3)-symmetric MPS cannot be invariant under
one-site translation (a translation-invariant one exists only as a non-clustering superposition of three translates),
so finite-$D$ SU(3)-symmetric states always carry some trimerization and show no such jumps. Phase boundaries in SU($N$) chains with the fundamental
representation on each site should therefore be located by finite-entanglement scaling, as done here, or by
spectroscopy that tracks the sign change of the marginal coupling directly.

At large $\alpha$ the effective central charge lies below the SU(3)$_2$ value $16/5$ for all $\alpha=4$--$10$ and
decreases with bond dimension where the crossover length is shortest, as expected for the massless flow to
SU(3)$_1$.

Three questions remain open. Reaching the SU(3)$_1$ value $c=2$ requires correlation lengths beyond the crossover
length of the SU(3)$_2\to$SU(3)$_1$ flow, i.e.\ bond dimensions well above those used here, the more so at larger
$\alpha$. The position of $\alpha_{c2}$ above $3.5$ and a
possible transition slightly below $\alpha_{c1}=0.48$ are limited by the exponentially slow opening of the gap at
both boundaries; a level-spectroscopy analysis of the transfer matrix or of excitations resolved by irrep would
determine both boundaries independently of finite-entanglement scaling.

\begin{figure*}[!tbp]
\centering
\includegraphics[width=0.85\textwidth]{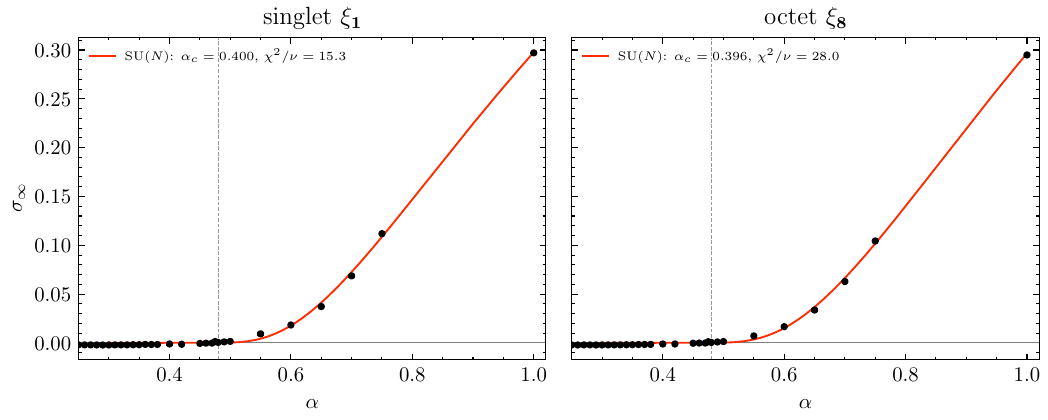}
\caption{Fit of the SU($N$) essential-singularity form, Eq.~(\ref{eq:esforms}), to the extrapolated order parameter, using the
singlet (left) and octet (right) correlation length in the extrapolation. The vertical line marks
$\alpha=0.48$. The plotted quantity is $\sigma_\infty=(\sqrt3/2)\,T_\infty$.}
\label{fig:fits}
\end{figure*}

\acknowledgements
The authors acknowledge support from the Simons Foundation in the framework of the call ``Centers of Excellence in Ukraine'', grant No.~SFI-PD-COE-00032628.

\appendix
\section{Essential-singularity fits}\label{app:fits}

\begin{figure*}[!tbp]
\centering
\includegraphics[width=0.85\textwidth]{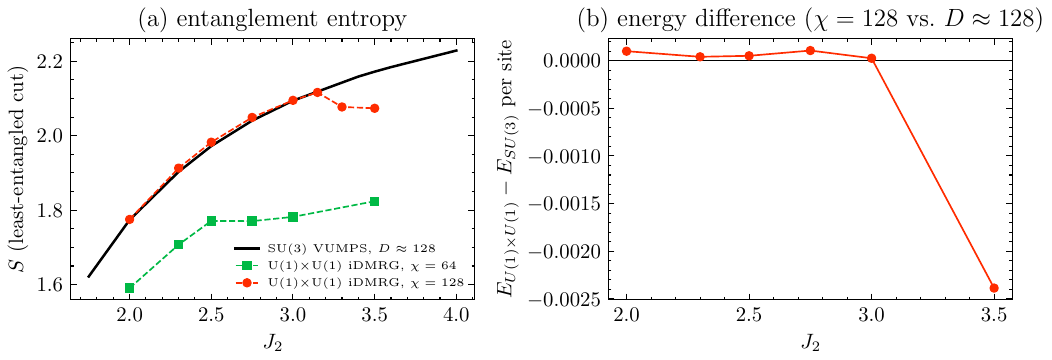}
\caption{(a) Entanglement entropy versus $J_2$ ($J_1 = 1$, so $J_2=\alpha$): SU(3)-symmetric VUMPS at $D\approx128$ (least-entangled cut) and
Abelian iDMRG (color numbers conserved modulo 6) at $\chi=64$ and $128$. (b) Energy difference per site between the
Abelian state at $\chi=128$ and the SU(3)-symmetric state with the nearest bond dimension ($D=129$--$141$).}
\label{fig:u1}
\end{figure*}

Figure~\ref{fig:fits} shows fits of $T_\infty(\alpha)$ on the window $\alpha\in[0.47,1.0]$. Here $T_\infty$ is
extrapolated from all available states of each coupling with $D\ge282$: up to $D\approx4000$ at $\alpha=0.47$--$0.50$,
$D\approx1000$--$1400$ at $\alpha=0.475$ and $0.55$--$0.75$, and $D\approx6000$ at $\alpha=1$. The window is fitted with
the essential-singularity form expected for the lower transition. At $\alpha_{c1}$, a marginal current–current coupling $g\propto\alpha-\alpha_c$ changes sign~\cite{HCL2023}. 
With SU($N$) symmetry, it is the only coupling that flows $dg/dl\propto g^2$, so that strong coupling is reached after $l^*\propto1/(\alpha-\alpha_c)$ and $\xi\propto e^{l^*}=\exp[b/(\alpha-\alpha_c)]$. 
Since $T_\infty\propto\xi^{-2/3}$,
\begin{equation}
  T_\infty=A\,e^{-b'/(\alpha-\alpha_c)}
  \label{eq:esforms}
\end{equation}
for $\alpha>\alpha_c$. With $T_\infty$ extrapolated using the singlet or the octet correlation length, the fit gives
$\alpha_c=0.400$ and $0.396$ ($\pm0.007$ and $\pm0.005$), with $\chi^2/\nu=15$ and $28$; on the narrower windows
$\alpha\in[0.50,1.0]$ and $[0.55,1.0]$ it gives $\alpha_c=0.399$--$0.406$, with $\chi^2/\nu$ up to $58$. Adding a constant
offset as a free parameter (to absorb the bias discussed in Sec.~\ref{sec:lower}) moves $\alpha_c$ to $0.40$--$0.49$. The location of the
transition cannot be obtained in this way at reachable bond dimensions.

\section{Abelian iDMRG check}\label{app:u1}

\begin{figure*}[!tbp]
\centering
\includegraphics[width=\textwidth]{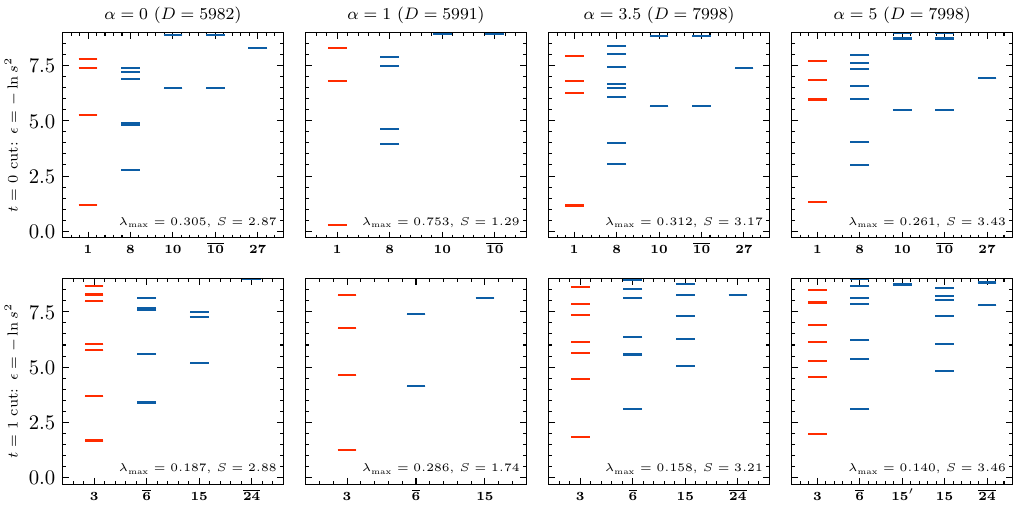}
\caption{Entanglement spectra resolved by SU(3) irrep on the two inequivalent cuts (triality $t=0,1$; the $t=2$
cut is the conjugate of $t=1$), at the Sutherland point ($\alpha=0$), in the trimerized phase ($\alpha=1$), near
the upper boundary ($\alpha=3.5$) and at large $\alpha$ ($\alpha=5$). Each level of irrep $\lambda$ is
$\dim\lambda$-fold degenerate; $\mathbf{15}'$ denotes the irrep with Dynkin labels $(4,0)$.}
\label{fig:es}
\end{figure*}

\begin{table}[!htbp]
\caption{Abelian iDMRG states compared with SU(3)-symmetric VUMPS states with the same number of states: energy
difference per site $\Delta E=E_{\rm Ab}-E_{\rm SU(3)}$ (no SU(3)-symmetric state with $D<46$, so none at $\chi=32$),
color-permutation asymmetry $A$ of the Schmidt weights, and spread $\sigma_{\mathbf 8}$ of the octet weights. Both
$A$ and $\sigma_{\mathbf 8}$ vanish for an SU(3)-symmetric state. The jump at $\chi=128$ lies between $J_2=3.15$ and
$3.3$.}
\label{tab:u1}
\begin{ruledtabular}
\begin{tabular}{ccrcc}
$J_2$ & $\chi$ & $\Delta E$ & $A$ & $\sigma_{\mathbf 8}$\\
\hline
2.0  & 32  & ---                     & 0.009 & 0.41\\
2.0  & 64  & $-3.5\times10^{-4}$     & 0.015 & 0.14\\
2.0  & 128 & $+2.8\times10^{-5}$     & 0.002 & 0.02\\
3.0  & 32  & ---                     & 0.286 & 2.43\\
3.0  & 64  & $-1.9\times10^{-3}$     & 0.034 & 0.80\\
3.0  & 128 & $-3.0\times10^{-5}$     & 0.018 & 0.48\\
3.15 & 32  & ---                     & 0.285 & 2.43\\
3.15 & 64  & $-2.3\times10^{-3}$     & 0.039 & 0.87\\
3.15 & 128 & $-1.1\times10^{-4}$     & 0.022 & 0.60\\
3.5  & 32  & ---                     & 0.281 & 2.44\\
3.5  & 64  & $-3.6\times10^{-3}$     & 0.047 & 0.99\\
3.5  & 128 & $-2.5\times10^{-3}$     & 0.371 & 1.98\\
\end{tabular}
\end{ruledtabular}
\end{table}

The Abelian calculations use an independent infinite-DMRG code (ITensors~\cite{ITensor2022}) with the same six-site cell and Hamiltonian
convention as the VUMPS calculations, and final bond dimensions $\chi=64$ and $128$. The code conserves the numbers of
particles of two of the three colors modulo 6, i.e.\ a $\mathbb Z_6\times\mathbb Z_6$ subgroup of the
U(1)$\times$U(1) color symmetry used in Ref.~\cite{HCL2023}, which keeps the number of charge sectors on the bonds of
the infinite chain bounded; like Ref.~\cite{HCL2023}, it does not impose SU(3). Its local
bond energies are evaluated on a finite block with approximate environments and do not show the period-3 structure
of the infinite state; we therefore compare only the energy per site and the entanglement entropy, which the code
reports for one cut. The SU(3)-symmetric reference states for this comparison were computed at small bond dimension
($D=46$, $63$, $90$, $123$, $160$) by two-site growth and VUMPS, as in the main calculations; their energies and
least-entangled-cut entropies are interpolated linearly in $\ln D$ to $D=\chi$. At $\chi=128$ the Abelian and the
SU(3)-symmetric energies agree within $1.1\times10^{-4}$ per site for $J_2=2.0$--$3.15$, and the entropies within
$0.01$. The entropy of the Abelian state rises smoothly up to $J_2=3.15$ ($S=2.116$) and drops to $2.077$ at $J_2=3.3$
and $2.074$ at $3.5$, where the SU(3)-symmetric state has $S=2.167$ (Fig.~\ref{fig:u1}).

To test the symmetry of the Abelian states we record, at the end of each $\chi$ stage, the Schmidt values $s_k$ of
the central cut together with the color charges $(N_1,N_2)\bmod 6$ of the Schmidt states. Taking the charges relative
to the largest Schmidt state, with $\Delta N_3=-\Delta N_1-\Delta N_2$, we form the summed weight $P(w)$ of the Schmidt
states with color weight $w=(\Delta N_1,\Delta N_2,\Delta N_3)$ and the asymmetry
\begin{equation}
  A=\max_{\pi\in S_3}\tfrac12\sum_w\big|P(w)-P(\pi w)\big| ,
\end{equation}
which vanishes for an SU(3)-symmetric state. The test is necessary but not sufficient, since a splitting inside a
multiplet that keeps the color-permutation symmetry leaves $A=0$; we therefore also record the relative spread
$\sigma_{\mathbf 8}=(\max-\min)/{\rm mean}$ of the eight weights $s_2^2,\dots,s_9^2$, which in an SU(3)-symmetric state
form the octet following the singlet. Table~\ref{tab:u1} lists both together with the energy difference. Before the
jump both measures decrease with $\chi$; at $J_2=3.5$ and $\chi=128$ the state is strongly SU(3)-breaking. Repeated
runs of the code at $\chi=64$ reach different broken states, with energies differing by up to $5.5\times10^{-3}$ per
site, while the $\chi=128$ results are reproduced exactly; the table uses one set of runs throughout.

\section{No one-site translation invariance of injective SU(3)-symmetric MPS}\label{app:proof}

\begin{figure}[!htbp]
\centering
\includegraphics[width=\columnwidth]{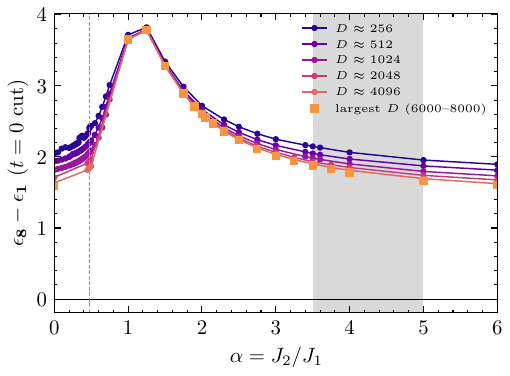}
\caption{Singlet--octet entanglement gap $\epsilon_{\mathbf 8}-\epsilon_{\mathbf 1}$ on the $t=0$ cut, at bond
dimensions matched to Ref.~\cite{HCL2023} and at the largest $D$. Dashed line: $\alpha_{c1}=0.48$; grey band: $\alpha=3.5$--$5$,
where the upper boundary lies (Sec.~\ref{sec:upper}).}
\label{fig:esgap}
\end{figure}

Consider a uniform MPS with tensors $A^s$, $s=1,2,3$, acting on a virtual space $V$ of dimension $D$, and its
transfer map $\mathcal E(X)=\sum_sA^sXA^{s\dagger}$. The MPS is injective (normal) if, after normalization, $1$ is
the only eigenvalue of $\mathcal E$ of modulus one, it is non-degenerate, and its eigenvector $\rho$ is positive
definite; equivalently, connected correlations decay exponentially and the state is the unique ground state of a
local parent Hamiltonian. We show that such an MPS cannot be invariant under SU(3) acting in the fundamental
representation $u(g)$ on every site.

(i) If an injective MPS is invariant under $u(g)^{\otimes N}$ for all $N$, there are a unitary representation $V_g$
on $V$, defined up to phases, and a one-dimensional representation $e^{i\theta_g}$ such that
$\sum_{s'}u(g)_{ss'}A^{s'}=e^{i\theta_g}V_gA^sV_g^{\dagger}$~\cite{PerezGarcia2008,LiebSchultzMattis1961, AffleckLieb1986,Sanz2009}.

(ii) SU(3) has no non-trivial one-dimensional representations, so $e^{i\theta_g}=1$; it is simply connected, so
$V_g$ can be chosen as a linear representation and decomposes into irreps. By Schur's lemma the center element
$z=\omega\one$, $\omega=e^{2\pi i/3}$, acts on an irrep of triality $t$ as $\omega^t$, so that
$V_z=Z\equiv\bigoplus_t\omega^t\,\one_{V_t}$, where $V_t$ collects the irreps of triality $t$.

(iii) In the fundamental representation $u(z)=\omega$. Hence $ZA^sZ^{-1}=\omega A^s$: $A^s$ maps $V_t$ to $V_{t+1}$.

(iv) Using $A^sZ=\omega^{-1}ZA^s$, $\mathcal E(Z\rho)=\sum_sA^sZ\rho A^{s\dagger}=\omega^{-1}Z\mathcal E(\rho)=\omega^{-1}Z\rho$.
Since $Z$ is invertible and $\rho$ is positive definite, $Z\rho\neq0$; thus $\omega^{-1}\neq1$ is a second
eigenvalue of $\mathcal E$ of modulus one, which contradicts injectivity.

The same argument applied to the tensor of an $L$-site cell, on which the center acts as $\omega^L$, shows that an
injective SU(3)-symmetric MPS requires $L\equiv0\pmod3$; the six-site cell used here satisfies this. If the
injectivity assumption is dropped, a one-site invariant SU(3)-symmetric MPS exists: $V=V_0\oplus V_1\oplus V_2$ with
$A^s$ cyclic between the blocks, for which $\mathcal E$ has the eigenvalues $1$, $\omega$ and $\omega^2$ on the unit
circle. In the thermodynamic limit this state is the equal-weight superposition of the three translates of a
period-3 injective state; its correlations do not cluster. The argument is the SU(3) counterpart of the one for SU(2)
with half-integer spin, where the center $\mathbb Z_2$ plays the role of $\mathbb Z_3$~\cite{LiebSchultzMattis1961, AffleckLieb1986,Sanz2009}.

\section{Entanglement spectra}\label{app:es}

Figure~\ref{fig:es} shows the entanglement spectra, $\epsilon=-\ln\lambda$ with $\lambda$ the eigenvalues of the
half-chain reduced density matrix, resolved by SU(3) irrep. At $\alpha=1$ the $t=0$ cut is dominated by a single
singlet level ($\lambda_{\max}=0.75$), separated by $\Delta\epsilon=3.7$ from the first octet, and the neighboring cuts
by a single $\mathbf 3$ or $\bar{\mathbf 3}$, as for a product of SU(3) singlets on three sites. At $\alpha=0$, $3.5$
and $5$ the spectra are dense, with many irreps at low entanglement energy. Figure~\ref{fig:esgap} shows the gap
$\epsilon_{\mathbf 8}-\epsilon_{\mathbf 1}$ between the lowest octet and the lowest singlet level on the $t=0$ cut
(cf.\ Fig.~4, bottom, of Ref.~\cite{HCL2023}). It peaks at $\alpha\approx1.25$, where it is independent of $D$, and
decreases slowly with $D$ outside the trimerized phase. It shows no jump at any $D$.

\section*{Code availability}
The code used to obtain and analyze the data (based on MPSKit and TensorKit~\cite{MPSKit,TensorKit2025}) will be made
publicly available upon publication.

\FloatBarrier
\bibliography{references_zigzag}

@article{Sutherland1975,
  author  = {Sutherland, Bill},
  title   = {Model for a multicomponent quantum system},
  journal = {Phys. Rev. B},
  volume  = {12},
  pages   = {3795--3805},
  year    = {1975},
  doi     = {10.1103/PhysRevB.12.3795}
}

@article{GorshkovNatPhys2010,
  author  = {Gorshkov, A. V. and Hermele, M. and Gurarie, V. and Xu, C. and Julienne, P. S. and Ye, J. and Zoller, P. and Demler, E. and Lukin, M. D. and Rey, A. M.},
  title   = {Two-orbital {SU(N)} magnetism with ultracold alkaline-earth atoms},
  journal = {Nat. Phys.},
  volume  = {6},
  pages   = {289--295},
  year    = {2010},
  doi     = {10.1038/nphys1535}
}

@article{CazalillaRey2014,
  author  = {Cazalilla, M. A. and Rey, A. M.},
  title   = {Ultracold {Fermi} gases with emergent {SU(N)} symmetry},
  journal = {Rep. Prog. Phys.},
  volume  = {77},
  pages   = {124401},
  year    = {2014},
  doi     = {10.1088/0034-4885/77/12/124401}
}

@article{Taie2012,
  author  = {Taie, S. and Yamazaki, R. and Sugawa, S. and Takahashi, Y.},
  title   = {An {SU(6)} {Mott} insulator of an atomic {Fermi} gas realized by large-spin {Pomeranchuk} cooling},
  journal = {Nat. Phys.},
  volume  = {8},
  pages   = {825--830},
  year    = {2012},
  doi     = {10.1038/nphys2430}
}

@article{Pagano2014,
  author  = {Pagano, G. and Mancini, M. and Cappellini, G. and Lombardi, P. and Sch{\"a}fer, F. and Hu, H. and Liu, X.-J. and Catani, J. and Sias, C. and Inguscio, M. and Fallani, L.},
  title   = {A one-dimensional liquid of fermions with tunable spin},
  journal = {Nat. Phys.},
  volume  = {10},
  pages   = {198--201},
  year    = {2014},
  doi     = {10.1038/nphys2878}
}

@article{Zhang2014,
  author  = {Zhang, X. and Bishof, M. and Bromley, S. L. and Kraus, C. V. and Safronova, M. S. and Zoller, P. and Rey, A. M. and Ye, J.},
  title   = {Spectroscopic observation of {SU(N)}-symmetric interactions in {Sr} orbital magnetism},
  journal = {Science},
  volume  = {345},
  pages   = {1467--1473},
  year    = {2014},
  doi     = {10.1126/science.1254978}
}

@article{Corboz2007trimer,
  author  = {Corboz, Philippe and L{\"a}uchli, Andreas M. and Totsuka, Keisuke and Tsunetsugu, Hirokazu},
  title   = {Spontaneous trimerization in a bilinear-biquadratic {S=1} zig-zag chain},
  journal = {Phys. Rev. B},
  volume  = {76},
  pages   = {220404},
  year    = {2007},
  doi     = {10.1103/PhysRevB.76.220404}
}

@article{ITensor2022,
  author  = {Fishman, Matthew and White, Steven R. and Stoudenmire, E. Miles},
  title   = {The {ITensor} Software Library for Tensor Network Calculations},
  journal = {SciPost Phys. Codebases},
  pages   = {4},
  year    = {2022},
  doi     = {10.21468/SciPostPhysCodeb.4}
}

@article{Calabrese2004,
  author  = {Calabrese, Pasquale and Cardy, John},
  title   = {Entanglement entropy and quantum field theory},
  journal = {J. Stat. Mech.},
  pages   = {P06002},
  year    = {2004},
  doi     = {10.1088/1742-5468/2004/06/P06002}
}

@article{Pollmann2009,
  author  = {Pollmann, Frank and Mukerjee, Subroto and Turner, Ari M. and Moore, Joel E.},
  title   = {Theory of finite-entanglement scaling at one-dimensional quantum critical points},
  journal = {Phys. Rev. Lett.},
  volume  = {102},
  pages   = {255701},
  year    = {2009},
  doi     = {10.1103/PhysRevLett.102.255701}
}

@article{Nataf2016,
  author  = {Nataf, Pierre and Mila, Fr{\'e}d{\'e}ric},
  title   = {Exact diagonalization of {Heisenberg} {SU(N)} chains in the fully symmetric and antisymmetric representations},
  journal = {Phys. Rev. B},
  volume  = {93},
  pages   = {155134},
  year    = {2016},
  doi     = {10.1103/PhysRevB.93.155134},
  eprint  = {1601.00922}
}

@article{Hofrichter2016,
  author  = {Hofrichter, Christian and Riegger, Luis and Scazza, Francesco and H{\"o}fer, Moritz and Fernandes, Diogo Rio and Bloch, Immanuel and F{\"o}lling, Simon},
  title   = {Direct probing of the {Mott} crossover in the {SU(N)} {Fermi-Hubbard} model},
  journal = {Phys. Rev. X},
  volume  = {6},
  pages   = {021030},
  year    = {2016},
  doi     = {10.1103/PhysRevX.6.021030}
}

@article{ZaunerStauber2018,
  author  = {Zauner-Stauber, V. and Vanderstraeten, L. and Fishman, M. T. and Verstraete, F. and Haegeman, J.},
  title   = {Variational optimization algorithms for uniform matrix product states},
  journal = {Phys. Rev. B},
  volume  = {97},
  pages   = {045145},
  year    = {2018},
  doi     = {10.1103/PhysRevB.97.045145},
  eprint  = {1701.07035}
}

@article{Vanderstraeten2019,
  author  = {Vanderstraeten, Laurens and Haegeman, Jutho and Verstraete, Frank},
  title   = {Tangent-space methods for uniform matrix product states},
  journal = {SciPost Phys. Lect. Notes},
  pages   = {7},
  year    = {2019},
  doi     = {10.21468/SciPostPhysLectNotes.7},
  eprint  = {1810.07006}
}

@article{HCL2023,
  author  = {Herviou, Lo{\"i}c and Capponi, Sylvain and Lecheminant, Philippe},
  title   = {Even-odd effects in the {$J_1$-$J_2$} {SU($N$)} {Heisenberg} spin chain},
  journal = {Phys. Rev. B},
  volume  = {107},
  pages   = {205135},
  year    = {2023},
  doi     = {10.1103/PhysRevB.107.205135},
  eprint  = {2302.14090},
  archivePrefix = {arXiv}
}

@article{Rachel2009,
  author  = {Rachel, Stephan and Thomale, Ronny and F{\"u}hringer, Max and Schmitteckert, Peter and Greiter, Martin},
  title   = {Spinon confinement and the {Haldane} gap in {SU($n$)} spin chains},
  journal = {Phys. Rev. B}, volume = {80}, pages = {180420(R)}, year = {2009},
  doi     = {10.1103/PhysRevB.80.180420}, eprint = {0904.3882}, archivePrefix = {arXiv}
}

@article{NatafGozelMila2021,
  author  = {Nataf, Pierre and Gozel, Samuel and Mila, Fr{\'e}d{\'e}ric},
  title   = {Edge states and universality class of the critical two-box symmetric {SU(3)} chain},
  journal = {Phys. Rev. B}, volume = {104}, pages = {L180411}, year = {2021},
  doi     = {10.1103/PhysRevB.104.L180411}, eprint = {2107.09306}, archivePrefix = {arXiv}
}

@article{Sanz2009,
  author  = {Sanz, M. and Wolf, M. M. and P{\'e}rez-Garc{\'i}a, D. and Cirac, J. I.},
  title   = {Matrix product states: Symmetries and two-body {Hamiltonians}},
  journal = {Phys. Rev. A}, volume = {79}, pages = {042308}, year = {2009},
  doi     = {10.1103/PhysRevA.79.042308}, eprint = {0901.2223}, archivePrefix = {arXiv}
}

@article{Eggert1996,
  author  = {Eggert, Sebastian},
  title   = {Numerical evidence for multiplicative logarithmic corrections from marginal operators},
  journal = {Phys. Rev. B}, volume = {54}, pages = {R9612}, year = {1996},
  doi     = {10.1103/PhysRevB.54.R9612}, eprint = {cond-mat/9602026}, archivePrefix = {arXiv}
}

@misc{TensorKit2025,
  author  = {Devos, Lukas and Haegeman, Jutho},
  title   = {{TensorKit.jl}: A {Julia} package for large-scale tensor computations, with a hint of category theory},
  year    = {2025}, eprint = {2508.10076}, archivePrefix = {arXiv}, doi = {10.48550/arXiv.2508.10076}
}

@misc{MPSKit,
  author  = {Devos, Lukas and Van Damme, Maarten and Haegeman, Jutho},
  title   = {{MPSKit.jl}}, note = {version 0.13.13},
  doi     = {10.5281/zenodo.10654900}, url = {https://github.com/QuantumKitHub/MPSKit.jl}, year = {2026}
}

@article{Alex2011,
  author  = {Alex, Arne and Kalus, Matthias and Huckleberry, Alan and von Delft, Jan},
  title   = {A numerical algorithm for the explicit calculation of {SU($N$)} and {SL($N$,$\mathbb{C}$)} {Clebsch-Gordan} coefficients},
  journal = {J. Math. Phys.}, volume = {52}, pages = {023507}, year = {2011},
  doi     = {10.1063/1.3521562}, eprint = {1009.0437}, archivePrefix = {arXiv}
}

@misc{SUNRepresentations,
  title   = {{SUNRepresentations.jl}}, howpublished = {\url{https://github.com/QuantumKitHub/SUNRepresentations.jl}},
  note    = {accessed October 2026}, year = {2026}
}

@article{PerezGarcia2008,
  author  = {P{\'e}rez-Garc{\'i}a, D. and Wolf, M. M. and Sanz, M. and Verstraete, F. and Cirac, J. I.},
  title   = {String order and symmetries in quantum spin lattices},
  journal = {Phys. Rev. Lett.},
  volume  = {100},
  pages   = {167202},
  year    = {2008},
  doi     = {10.1103/PhysRevLett.100.167202},
  eprint  = {0802.0447}
}

@article{Weichselbaum2012,
  author  = {Weichselbaum, Andreas},
  title   = {Non-abelian symmetries in tensor networks: A quantum symmetry space approach},
  journal = {Ann. Phys.},
  volume  = {327},
  pages   = {2972--3047},
  year    = {2012},
  doi     = {10.1016/j.aop.2012.07.009},
  eprint  = {1202.5664}
}

@article{McCullochGulacsi2002,
  author  = {McCulloch, I. P. and Gul{\'a}csi, M.},
  title   = {The non-{Abelian} density matrix renormalization group algorithm},
  journal = {Europhys. Lett.},
  volume  = {57},
  pages   = {852},
  year    = {2002},
  doi     = {10.1209/epl/i2002-00393-0},
  eprint  = {cond-mat/0012319}
}

@article{Tagliacozzo2008,
  title = {Scaling of entanglement support for matrix product states},
  author = {Tagliacozzo, L. and de Oliveira, Thiago. R. and Iblisdir, S. and Latorre, J. I.},
  journal = {Phys. Rev. B},
  volume = {78},
  issue = {2},
  pages = {024410},
  numpages = {14},
  year = {2008},
  month = {Jul},
  publisher = {American Physical Society},
  doi = {10.1103/PhysRevB.78.024410},
  url = {https://link.aps.org/doi/10.1103/PhysRevB.78.024410}
}

@article{Pirvu2012,
  title = {Matrix product states for critical spin chains: Finite-size versus finite-entanglement scaling},
  author = {Pirvu, B. and Vidal, G. and Verstraete, F. and Tagliacozzo, L.},
  journal = {Phys. Rev. B},
  volume = {86},
  issue = {7},
  pages = {075117},
  numpages = {16},
  year = {2012},
  month = {Aug},
  publisher = {American Physical Society},
  doi = {10.1103/PhysRevB.86.075117}
}

@article{Affleck1986,
  author  = {Affleck, Ian},
  title   = {Exact critical exponents for quantum spin chains, non-linear $\sigma$-models at $\theta=\pi$ and the quantum {Hall} effect},
  journal = {Nucl. Phys. B},
  volume  = {265},
  pages   = {409},
  year    = {1986}
}

@article{Affleck1988,
  author  = {Affleck, Ian},
  title   = {Critical behaviour of {SU($n$)} quantum chains and topological non-linear $\sigma$-models},
  journal = {Nucl. Phys. B},
  volume  = {305},
  pages   = {582},
  year    = {1988}
}

@article{Coleman1973,
  author  = {Coleman, Sidney},
  title   = {There are no {G}oldstone bosons in two dimensions},
  journal = {Commun. Math. Phys.},
  volume  = {31},
  number  = {4},
  pages   = {259--264},
  year    = {1973},
  doi     = {10.1007/BF01646487}
}

@article{LiebSchultzMattis1961,
  author  = {Lieb, Elliott and Schultz, Theodore and Mattis, Daniel},
  title   = {Two soluble models of an antiferromagnetic chain},
  journal = {Ann. Phys.},
  volume  = {16},
  number  = {3},
  pages   = {407--466},
  year    = {1961},
  doi     = {10.1016/0003-4916(61)90115-4}
}

@article{AffleckLieb1986,
  author  = {Affleck, Ian and Lieb, Elliott H.},
  title   = {A proof of part of {H}aldane's conjecture on spin chains},
  journal = {Lett. Math. Phys.},
  volume  = {12},
  number  = {1},
  pages   = {57--69},
  year    = {1986},
  doi     = {10.1007/BF00400304}
}

@article{Lecheminant2015,
  author        = {Lecheminant, P.},
  title         = {Massless renormalization group flow in {SU}({N})$_k$ perturbed conformal field theory},
  journal       = {Nucl. Phys. B},
  volume        = {901},
  pages         = {510--525},
  year          = {2015},
  doi           = {10.1016/j.nuclphysb.2015.11.004},
  eprint        = {1509.01680},
  archivePrefix = {arXiv}
}

@article{ItoiKato1997,
  author        = {Itoi, Chigak and Kato, Masa-Hide},
  title         = {Extended massless phase and the {H}aldane phase in a spin-1 isotropic antiferromagnetic chain},
  journal       = {Phys. Rev. B},
  volume        = {55},
  number        = {13},
  pages         = {8295--8303},
  year          = {1997},
  doi           = {10.1103/PhysRevB.55.8295},
  eprint        = {cond-mat/9605105},
  archivePrefix = {arXiv}
}

@article{AffleckGepnerSchulzZiman1989,
  author  = {Affleck, I. and Gepner, D. and Schulz, H. J. and Ziman, T.},
  title   = {Critical behaviour of spin-$s$ {H}eisenberg antiferromagnetic chains: analytic and numerical results},
  journal = {J. Phys. A: Math. Gen.},
  volume  = {22},
  number  = {5},
  pages   = {511--529},
  year    = {1989},
  doi     = {10.1088/0305-4470/22/5/015}
}

\end{document}